\documentclass[aps,preprint,tighten,floats,epsf,rotate,showpacs]{revtex4}
\usepackage{graphicx}

\begin{document}

\title{Viscous instabilities in flowing foams: A Cellular Potts Model approach}
\author{Soma Sanyal \footnote{e-mail: ssanyal@indiana.edu.\\
 Presently at the School of Library and Information Science, Indiana University.}}
\author{James A. Glazier \footnote{e-mail: glazier@indiana.edu}}
\affiliation{Biocomplexity Institute, Department of Physics, Indiana University,
727, E. Third Street, Swain Hall West,  
 Bloomington, IN 47405-7105}

\begin{abstract}

The Cellular Potts Model ({\it CPM})  succesfully simulates  drainage
and shear in foams. Here we use the CPM to investigate
instabilities due to the flow of a single large bubble in a dry, monodisperse
two-dimensional flowing foam.  As in experiments in a Hele-Shaw cell, above a threshold 
velocity the large bubble
moves faster than the mean flow.  Our simulations reproduce analytical and experimental 
predictions for the velocity threshold and the relative velocity of the large bubble, 
demonstrating the utility of the CPM in foam rheology studies.

\vskip 0.5cm
\noindent Key words: {Foams,  Rheology, Bubbles and drops}
\pacs{ 83.80.Iz, 83.50.-v, 05.50.+q, 47.50.+d}

\end{abstract}

\maketitle

\section{Introduction}

 Foams' unusual rheology suits them to applications as diverse as efficient fire 
 supression and oil extraction in the petroleum industry \cite{eor,appli1}. Foams are 
 non-Newtonian, so understanding  their flow helps explain  the complex behaviours of 
 other structured fluids, which are difficult to investigate analytically. 
 Though we know that foams behave like solids under small stress and flow like  
 fluids under large stress, we  do not understand the relationship 
 between the macroscopic and microscopic properties of foams. 
 We still need experiments and  
 simulations to provide insight into foam-flow behaviour \cite{book,review}.
 Here, we show that the Cellular Potts Model ({\it CPM}) can successfully model dry, 
 {\it i.e} low fluid fraction, foam flow in a quasi-two dimensional {\it(2D)} Hele-Shaw 
({\it H-S}) cell, in which a single bubble layer flows lengthwise between two 
closely-spaced long and narrow parallel plates.  H-S flow is important in industry,
{\it e.g} in injection molding \cite{appli2} and display device manufacture \cite{appli3}.

The sizes and shapes of the bubbles in a foam may  
change due to gas diffusion between neighbouring bubbles, bubble coalescence, shear and 
drainage of the liquid in the walls between bubbles\cite{gopal}. This paper considers 
only shear-induced topological rearrangements or T1 
processes, where two bubbles come together to form a side, pushing apart two previously 
adjacent bubbles \cite{T1}. Since the timescale of the approach to a T1 depends on the
shear rate, while the usually fast relaxation time depends on the fluid-surface effective 
drag,at low shear rates bubble motion appears jagged. Under certain circumstances  
many  T1s occur together, each  T1 triggering the next,  forming 
an avalanche. Jiang {\it et. al.} \cite{jiang} have shown that the flow becomes  
smoother as the 
strain rate increases. However even at low strain rates, because the typical jump
size is a fraction of the size of a bubble, large aspect ratio flows, where bubbles 
are very small, appear smooth. The collective phenomena of foam
flow have inspired  many models, including constitutive, vertex, center, bubble 
and CPM models (\cite{consti}-\cite{potts}).
For example, Okuzano and Kawasaki used a vertex model to study the 
 effect of low shear rates on foams \cite{okuzono} and found avalanche-like rearrangements. 
Durian's bubble model \cite{durian} gave similar predictions but Weaire's center model 
\cite{Weaire} suggested that avalanche-like rearrangements are only possible for wet foams. 
Jiang {\it et. al.} tried to reconcile the different model predictions  
and experiments using the CPM \cite{jiang} \cite{potts}. 
They demonstrated hysteresis and avalanche-like rearrangements 
in a 2D non-coarsening foam and found that the T1 dynamics 
depended sensitively  the foam's topology.
Because the CPM derives from an equilibrium model, 
we must establish  its suitability to describe dynamic phenomena. Here we show that it 
correctly reproduces the rather subtle experimental behaviour of the flow of a large bubble 
in a background of small bubbles, further validating the use of the CPM to 
simulate flowing foams.

In a H-S cell, a monodisperse foam ( {\it i.e.}, a foam  of bubbles of equal size),
under a uniform pressure gradient, exhibits  simple plug flow. However,  
a polydisperse  ({\it i.e.}, a foam made of bubbles of different sizes) foam's 
 flow becomes unstable above a critical velocity. The size distribution of the bubbles 
then controls the velocity field, with the larger bubbles moving faster than the 
smaller ones, as experiments by Lordereau \cite{lordereau} have shown.
Recently  Cantat and Delannay \cite{cantat} studied the phenomenon in more detail, both 
 experimentally and numerically. Their experiments used a dry soap 
 froth contained in a H-S cell, with newly produced bubbles maintaining
 a steady pressure gradient along the length of the cell. Their simulations used a  
 vertex model with periodic boundary conditions along the direction
 of flow. Their analytical predictions for the critical velocity at which 
 a single large bubble begins to move faster than the bulk flow in an otherwise 
monodisperse foam agree with their numerical and experimental results.

According to Cantat and Delanney \cite{cantat}, below the critical velocity, all the 
bubbles in the foam move with a velocity $v_{0} {\bf{u_x}}$ where ${\bf{u_x}}$ 
is a unit vector in the direction of flow. The viscous force per unit surface, 
averaged on the scale of a bubble, is: 
\begin{equation}
{\mathbf{F}}_{visc} = - {\eta v_{0} \over d} {\mathbf{ u_x}},
\end{equation}
\noindent where $\eta$ is the effective viscosity and $d$ is the diameter of the small bubbles. 
The large bubble induces a pressure deficit, $\delta f = n \eta v_{0} d {\mathbf{u_x}}$,
where {\it n} is the number of films that would be present 
across the large bubble in the x direction if it were filled with small bubbles. 
Assuming the friction deficit is concentrated at $r = r_0$, the force equation becomes, 
\begin{equation}
{\mathbf F}_{visc} = - {\mathbf\nabla} {\eta v_{0} x\over d} + \delta ({\mathbf r - r_0}) 
{\eta v_{0} D^2 
\over d}{\mathbf u_x}, 
\end{equation}
\noindent where $D$ is the diameter of the large bubble. As the large bubble moves, 
it distorts the small bubbles and changes their stress distribution. Combining the forces 
due to surface tension and viscosity, 
Cantat and Delanney \cite{cantat} obtained the equations of motion:
\begin{equation}
-{\mathbf\nabla} ({\eta v_{0} x\over d} + P ) +  \mu {\mathbf \nabla^2 X} = -\delta ({\mathbf r - r_0}) {\eta v_{0} D^2 \over d}{\mathbf u_x}, 
\end{equation}
\begin{equation}
 {\mathbf{\nabla} \bullet \mathbf{X}} = 0,
\end{equation}
where $P$ is the pressure field given by, 
\begin{equation}
P = -{\eta v_{0} x\over d} + {\eta v_{0} D^2 \over 2 \pi d } {{x - x_0} \over (\mathbf {r-r_0})^2}, 
\end{equation}
for the small bubbles. The last term on the R.H.S of equation (5) gives the pressure 
discontinuity for the large bubble at 
$\mathbf {r} = {D \over 2\mathbf{u_x}}$, which must counterbalance the stress. The force balance 
gives the critical velocity: 
\begin{equation}
v_{c} \sim {\gamma \over \eta D},
\end{equation}
where $\gamma$ is the surface tension. 
The critical velocity is directly proportional to the surface tension and inversely 
proportional to the diameter of the large bubble and the viscosity.
 
In this work we use the CPM to reproduce   
large-bubble migration. Our results agree with the results in ref. \cite{cantat}. 
While CPM simulations are computationally simple,
we are not able to predict the viscosity analytically from model parameters, though we can 
 obtain an effective viscosity and other viscoelastic information from our simulations. 
In this respect, CPM simulations resemble experiments, in which we also cannot predict 
the effective foam viscosity from the fluid component's viscosity and surface tension 
\cite{appli1}. The capillary 
number appears to relate the velocity of the foam to fluid viscosity and surface tension, 
but experiments have shown it is not sufficient to describe the dynamic regime of a 
flowing foam \cite{dollet}.
New experiments have investigated the dependance of mobility on various parameters 
(\cite{dollet2}-\cite{kern}), but more experiments and analysis are still required.

   \section{The CPM}

Jiang {\it et. al.} \cite{jiang} provide details on the use of the CPM to study foam 
rheology. The CPM is lattice-based, with each lattice point 
having an integer {\it spin}. Like spins form  
{\it bubbles} while boundaries between unlike spins correspond to soap films. 
 The CPM Hamiltonian  contains a surface-energy term corresponding to film 
surface tension and a term constraining bubble areas corresponding to the conservation 
of mass within each bubble. The area constraint allows bubble compression according to the 
ideal gas law and transmits forces between bubbles, which is essential in a rheological 
simulation. We prevent coarsening, since in experiments the slow coarsening of bubbles
during their brief residence in a H-S cell is unmeasurable \cite{earnshaw}.

The CPM Hamiltonian thus has two terms:

\begin{equation}
H=\sum_{\vec{i},\vec{j}}J (1-\delta_{\sigma_{\vec{i}} \sigma_{\vec{j}}})+ \lambda \sum_n (a_n - A_n)^2, 
\end{equation} 

\noindent where $J$ is the coupling strength between spins 
$\sigma_{\vec{i}} ,\sigma_{\vec{j}} $ at neighboring lattice sites $\vec{i}$ and $\vec{j}$ and $\lambda$ is 
the inverse of the compressibility of the gas. 
$\it{A_n}$ is the area of a 
bubble with no forces (including surface tension) acting on it, 
which we call the {\it target area},
while $\it{a_n}$ is the current area of the same bubble as it flows. The difference 
between the areas $(a_n - A_n)$ gives a bubble's pressure. The first term gives the 
total surface energy and the second term the pressure energy. The CPM spins evolve 
according to a Modified Metropolis algorithm \cite{jiang}. Each time step corresponds 
to a complete Monte Carlo Sweep ({\it MCS}) of the lattice. 

%%%%%%%%%%%%%%%%%%%%%%%%%%%%%%%%%%%%%%%%%%%%%%%%%%%%%%%%%%%%%%%%%
\begin{figure}[b]
\begin{center}
\includegraphics[height=60mm]{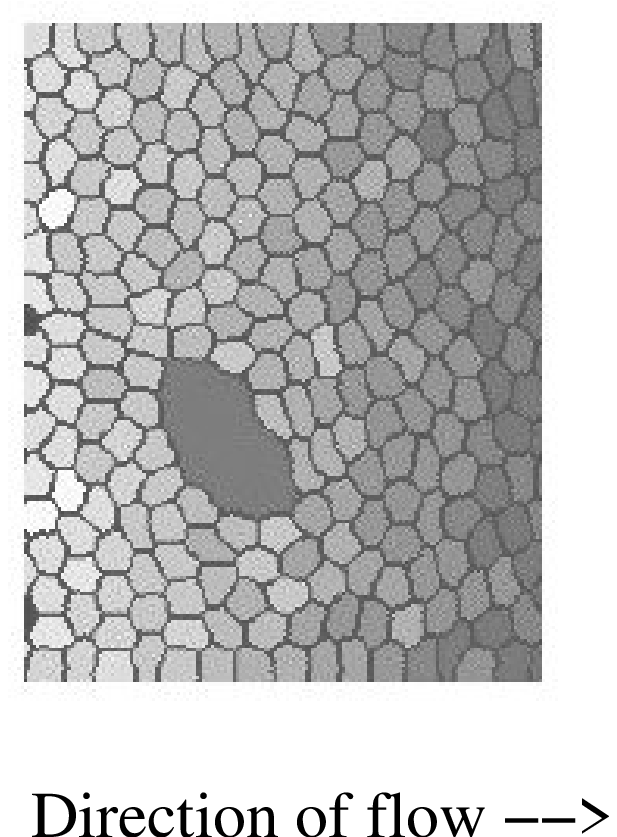}
\end{center}
\vskip -0.5cm
\caption{}{Detail of a CPM simulation of a 
 quasi-stationary flowing foam with a large bubble. 
 The shading denotes bubble pressures, with darker shades denoting lower pressures.}
\label{Fig.1}
\end{figure}
%%%%%%%%%%%%%%%%%%%%%%%%%%%%%%%%%%%%%%%%%%%%%%%%%%%%%%%%%%%%%%%%%%%%%
Fig. 1 shows a detail of a  simulation with a large bubble moving through  
smaller bubbles. The shades denote the pressure inside each bubble,  
darker shades denoting lower pressures. The bubbles move from left to right. Our lattice 
geometry is rectangular (usually 1000 X 200 sites) with open boundary conditions at the 
short sides, like a H-S experiment. 
We nucleate bubbles at a steady rate at one short end (the {\it head end}) and  remove 
them  at the opposite short end (the {\it tail end}).

All the bubbles, except the large bubble,  nucleate at a fraction of their target area 
(large pressure).  As they enter the lattice, they gradually expand, generating an excess 
pressure at the head end. As the bubbles move from left to right, they expand and their 
pressure decreases. When a bubble contacts the tail end of the lattice, we set its area 
constraint to zero so that it 
disappears  smoothly at near zero pressure.  Pressure differences between bubbles induce boundary movement
with a velocity proportional to the applied force \cite{jiang}.
This method of bubble creation and disappearance corresponds closely to the 
experiments  which generate bubbles continuously at one end of the channel and allow them 
to exit at near-atmospheric  pressure at the opposite end. Thus  
simulation and experiment both have a constant bubble-flux boundary condition at the 
head end and an absorbing boundary condition at the tail end.  
As we mentioned earlier, the absence of a simple relationship between the mobility of the 
bubbles and $J$ and $\lambda$ is a limitation of both the CPM and experiments.

Our simulations have $J=5 $ and $\lambda = 3$ and 
run at zero temperature.  The small bubbles target area is usually 625 lattice sites. 
We create a single large bubble of diameter $D$ at the first time step at a random 
position along the head end. We nucleate small bubbles  every 50 MCS with the initial 
sizes between 4 and 481 pixels. 
Varying the nucleation size of the bubbles at the head end 
changes the pressure gradient, which in turn changes the velocity of the flow.      
We also vary the large bubble size. If the large bubble radius is more than four times 
the small bubble radius, we use a larger lattice to avoid boundary effects.  
The small bubbles all have approximately the same velocity at any given time and we define 
the foam velocity as the average of the center-of-mass velocities of the
small bubbles at a fixed time. For each case, we run multiple replicas with different
random number generator seeds.
As in refs. \cite{okuzono} and \cite{jiang} we define the total
 {\it stored surface energy} $\phi$ to be, 
\begin{equation}
 \phi = \sum_{i,j}  (1 - \delta_{i,j}).
\end{equation}
The average stress tensor $\sigma$,  as  ref. \cite{okuzono} points out,  
relates directly  to  $\phi$. $\phi = Tr (\sigma)$. We scale out  differences due to 
initial conditions by using $\phi(t)\over\phi(0)$, where $\phi(0)$ is the value of $\phi$ 
at the start of the simulation. The applied strain rate is very high initially, then falls 
sharply to a low constant value, after which the applied strain is proportional to time 
and the energy {\it vs} time curve becomes equivalent to the energy {\it vs} applied 
strain curve. 
We call the flow  quasistationary when any drift in the total energy is 
 less than 2\% of the average energy over 1000 MCS and the bubble velocity  
changes by less than 10 \% of the average velocity over 1000 MCS.  We make all measurements in the
 quasistationary state.   Since bubbles we  introduce and eliminate  continuously  
we are never in a static state equilibrium. The finite H-S cell and pressure drop along it means that bubble
velocity varies down the cell length.

\section{Results}

%%%%%%%%%%%%%%%%%%%%%%%%%%%%%%%%%%%%%%%%%%%%%%%%%%%%%%%%%%%%%%%%%
\begin{figure}
\begin{center}
\includegraphics[width=80mm]{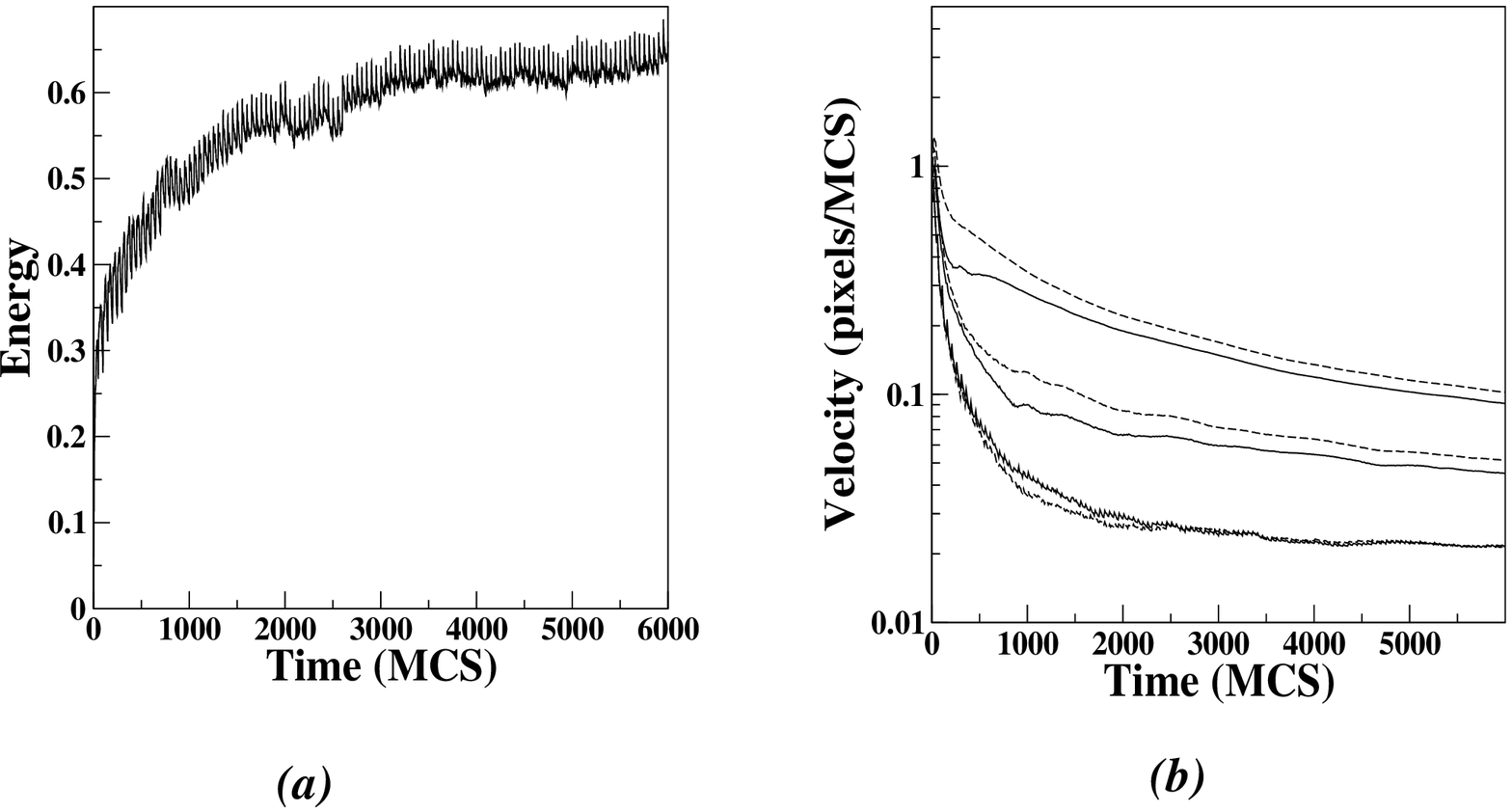}
\caption{}{(a) Equilibration of the stored surface energy of a simulated flowing foam 
containing a large bubble with small bubbles nucleating every 50 MCS. \\
(b) Equilibration of bubble velocities in a simulated flowing foam: 
large bubbledashed lines, small bubbles{\-}solid lines. 
The three pairs of curves are for different nucleation  
sizes of the small bubbles. The lowest curve corresponds to 
an initial nucleation size of 481 pixels and the higher curves to nucleation sizes of
156 and 25 pixels respectively.}
\label{Fig.2}
\vskip -0.5cm
\end{center}
\end{figure} 
%\caption{} }
%\label{Fig.2b}

%%%%%%%%%%%%%%%%%%%%%%%%%%%%%%%%%%%%%%%%%%%%%%%%%%%%%%%%%%%%%%%%%%%%%
Fig 2 (a)  plots  $\phi$ as a function of time and fig 2 (b) plots the velocity of large and 
small bubbles as a function of time. 
For slow flows, the small and large bubbles flow with the same velocity as a solid. Above a 
critical velocity the bubbles' velocities depend on their sizes, {\it e.g.}, our simulations the 
critical velocity is 0.014 pixels/MCS when the radius of the large 
bubble is twice that of the small bubbles.   
We plot the difference in velocity between the large bubble and the foam ($v_L-v_f$)
{\it vs} the velocity of the foam $v_f$ 
for different  large-bubble sizes $D$, scaling the velocity difference by $r = D/d$,
where the diameter of the small bubbles $d$ is  constant.  Fig. 3 shows our results.
 We checked that $v_c$  was independant of  $d$  by running  simulations 
with a small bubble target area, $A_n=400$ pixels.  We examined the cases $ r = 2, 3, 3.5$ and $4$.
Since very large bubbles tend to wobble and break into smaller bubbles, we analyzed only 
simulations in which  the large bubble traversed the H-S cell without breaking.
We find excellent agreement between our data and 
the theoretical form of Cantat and Delanney \cite{cantat}; 
\begin{equation}
{(v_L - v_f)\over v_f(D/d)}={-A \over v_f}\times{1\over Bln(1 - {A\over v_f})},
\end{equation} 
where $(v_L - v_f)$ is the difference between the 
large bubble velocity and the foam velocity and $A$ and $B$ are fitting parameters. The fitting value of
the critical velocity $A = 0.013$ pixels/MCS,   
$B = 8.66 $ is a dimensionless parameter which scales the 
velocity.  The asymptotic standard error for both parameters is less than $5\%$.

%%%%%%%%%%%%%%%%%%%%%%%%%%%%%%%%%%%%%%%%%%%%%%%%%%%%%%%%%%%%%%%%%
\begin{figure}
\begin{center}
\includegraphics[angle=270,width=80mm]{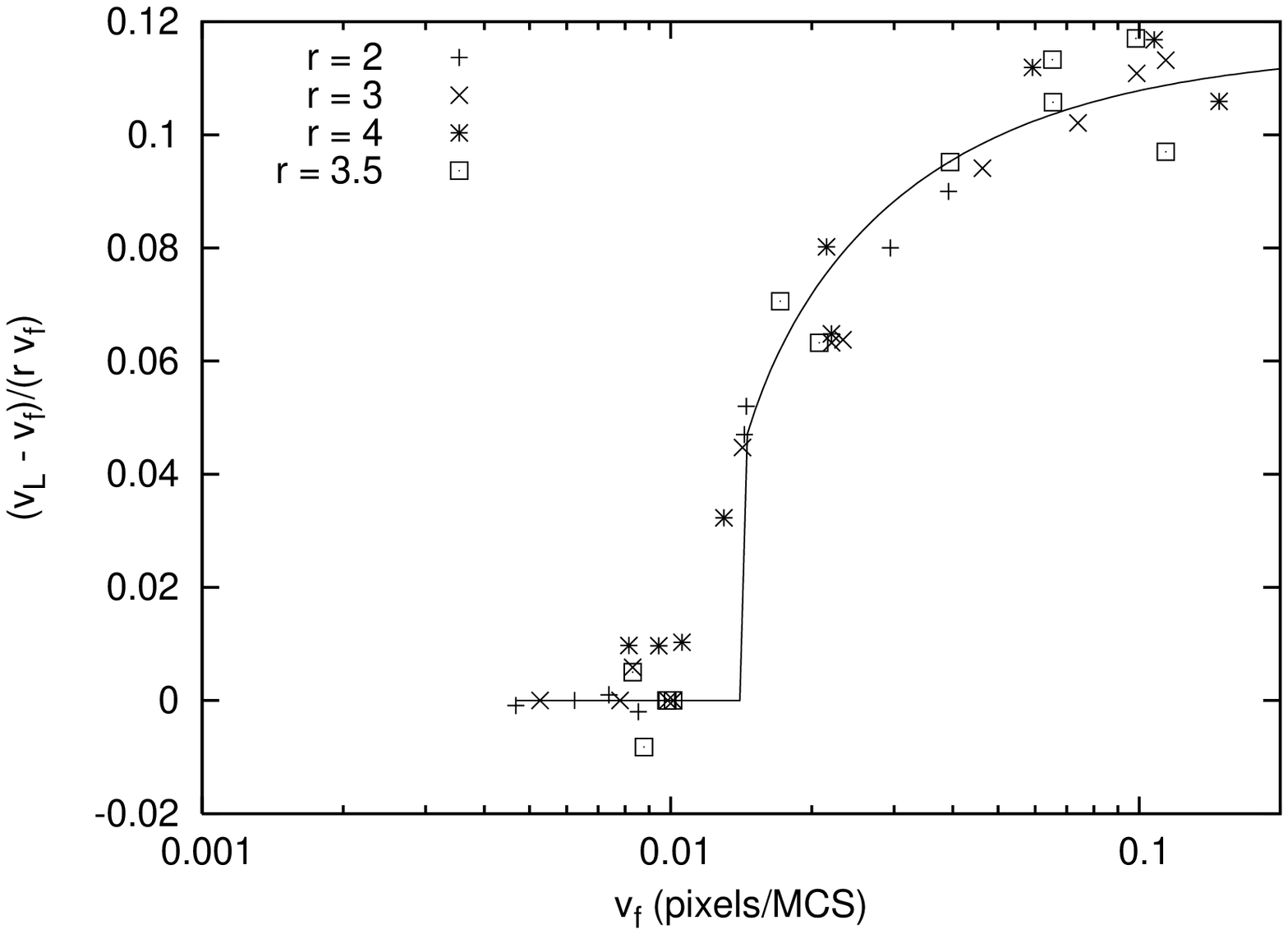}
\end{center}
\vskip -0.5cm
\caption{}{Difference between large bubble velocity $v_L$  and average foam 
 velocity, rescaled by $r v_f$, {\it vs} $v_f$ for different large-bubble sizes,
on a semilog scale. The symbols on the
graph correspond to different values of $r$. }
\label{Fig.4}
\end{figure}
%%%%%%%%%%%%%%%%%%%%%%%%%%%%%%%%%%%%%%%%%%%%%%%%%%%%%%%%%%%%%%%%%%%%%

The theoretical critical velocity is \cite{cantat}:
\begin{equation}
v_{c} = {\gamma h \over \eta D},
\end{equation}
where $h$ is the thickness of the H-S cell. Taking $h$ to be the small bubble size, 
we obtain $v_c$ from the values of $\gamma $ and $\eta$ obtained from our simulations. We obtain
$\eta$ by measuring  the relation between the effective pressure along the channel and the foam 
velocity and 
$\gamma$  from the size and pressure of the small bubbles. We find $v_c = 0.014$ pixels/MCS,
agreeing with the value which we obtained in the previous paragraph.

We can also calculate an analog of  the Deborah number ($N_D$) for our simulated bubble motion, 
the product of the shear strain rate and the event timescale 
\cite{durian}. The event timescale $\tau$  is the timescale of a T1, while the shear strain 
rate is the ratio of the velocity difference to the lengthscale (which in our case is the 
size of the small bubbles). So, 
 
\begin{equation} 
N_D = {(v_L - v_f) \tau \over d}.
\end{equation}

Our small bubble size is usually 25 pixels, while $\tau$ is approximately $20$ MCS so $N_D$ is 
between  $10^{-2}$ and $10^{-1}$ for most of our simulations. Our maximum $N_D$ is $0.08$.

  \section{Conclusions.}

We have shown that we can use the CPM to study the flow of a large bubble embedded 
in a monodisperse foam.  For small velocities, all the bubbles in the foam flow at 
the same velocity. Above a critical velocity, the velocities of the bubbles vary with their sizes. 
The critical velocity in our detailed simulations of a large bubble  
twice the size of the small bubbles matches very well with the critical velocity we obtain
 by fitting our simulation results for bubbles of various sizes to the analytical equation of 
Cantat and Delennay \cite{cantat},  and with the  
critical velocity we deduce theoretically from the effective viscosity and surface tension 
of the simulated foam. 
  
We have also checked that in a polydisperse foam,  
above a critical velocity different-size bubbles travel at different 
velocities, the large bubbles traveling faster. The dimensional form of the 
definition of the critical velocity suggests that bubbles of different sizes should have 
different critical velocities, however we have not been able 
to verify this dependance in a simulated  polydisperse foam because
the very large scatter in the velocity difference prevents us from identifying the critical 
velocities.  Experiments by Park and 
Durian revealed fingering instabilities in radial H-S cells \cite{park} which may relate to the 
viscous instability in rectangular H-S cells. However the aspect ratios
 of these experiments are very different from those in our simulations, so direct comparison 
is difficult.

{\bf Acknowledgements:} We acknowledge the use of AVIDD, the 
distributed computing facility at Indiana University Bloomington to run 
our simulations. We thank Debasis Dan, Ariel Balter, Lenhilson 
Coutinho, Roeland Merks and Julio Espinoza Ortiz for discussions, suggestions 
and comments. We also thank Isabelle Cantat for supplying the 
preprints which we cite in this paper and discussing the details 
of her experiments and simulations. We also acknowledge support from NSF grant IBN-0083653, 
NASA grant NAG2-1619, an IBM Innovation Institute award, an Indiana University Pervasive 
Technologies Laboratory Fellowship and the Biocomplexity Institute.

\end{document}